\RequirePackage{fix-cm}
\documentclass[smallextended]{svjour3}       
\smartqed  
\usepackage{graphicx}
\begin{document}

\title{The International Pulsar Timing Array:
}
\subtitle{A Galactic Scale Gravitational Wave Observatory}


\author{Maura McLaughlin
}


\institute{M. McLaughlin \at
	      West Virginia University\\
	      Department of Physics and Astronomy\\
              Morgantown WV 26506, USA\\
              Tel.: +1-304-293-4812\\
              Fax: +1-304-293-5732\\
              \email{maura.mclaughlin@mail.wvu.edu}           
}

\date{Accepted: 16 September 2014}

\maketitle

\begin{abstract}
The phenomenal rotational stability of millisecond pulsars  allows them to be used as
precise celestial clocks. An array of these pulsars can be exploited to search for correlated perturbations in their pulse
times of arrival due to
gravitational waves. Here, I describe the observations and analysis necessary to accomplish this goal and
present an overview of the
efforts of the worldwide pulsar timing community. Due to a growing number of millisecond pulsar discoveries,
improved instrumentation, and growing timespans of observation, the sensitivity of our pulsar timing array experiments is
expected to dramatically increase
over the next several years, leading to either a gravitational wave detection or very
 stringent constraints on low-frequency gravitational wave source populations before the end of the decade.
\keywords{pulsars \and gravitational waves \and general relativity \and black holes  \and radio astronomy}
\end{abstract}

\section{Introduction}
\label{sec:intro}

One of the fundamental predictions of Einstein's  theory of general relativity is the existence of 
gravitational waves (GWs), ripples in space-time produced by accelerating massive objects.
These waves travel at the speed of light, carry energy, and cause changes in the light travel time between objects.
We are confident of their existence from measurements of orbital decay due to energy loss from GW emission in double neutron
star binary systems
 \cite{tw82,ksm+06}. However,
as of yet, we have not detected the influence of GWs on space-time through a measured change in light travel time between two objects. This {\it direct} detection of GWs will
 allow us to robustly test the  predictions of General Relatively through the  measurement of GW properties,
providing spectacular proof of Einstein's theories. It will also usher in a new era of astrophysics in which we can use GWs
to study objects which are thus far invisible or
inaccessible through electromagnetic observations.

Pulsars are rapidly rotating, highly magnetized neutron stars produced in the supernova explosions of massive stars.
With the largest radio telescopes in the world, we can measure the arrival times of pulses
 from these objects very precisely. 
A PTA is a spatial array of pulsars  that is analyzed to detect correlated timing perturbations \cite{fb90}.
There are many uses for PTAs \cite{chm+10,hcm+12}, including searching 
for GW-induced perturbations with a characteristic quadrupolar angular correlation \cite{hd83}. The GWs that we are sensitive to will have low frequencies (of order $10^{-7} - 10^{-9}$~Hz),  complementary to the much higher
frequencies probed by ground- or space-based GW detectors. The most likely GW sources for detection by PTAs include supermassive black hole (SMBH) binaries \cite{sesana13},  cosmic strings \cite{Siemens:2006vk}, and, possibly, early universe phase transitions \cite{ccd+10} and inflation \cite{PhysRevD.37.2078,tong14}. Therefore, PTAs will provide crucial input to galaxy formation and evolution scenarios and cosmology \cite{sesana13_2}.

This article will 
 begin with a discussion of the properties of the detectors, rapidly rotating millisecond pulsars, or MSPs. It will then
 describe the experimental details of how one can use MSP timing measurements to detect GWs. It will review 
the properties of the world-wide MSP radio timing programs that are part of the International Pulsar Timing Array (IPTA).
Finally, it will conclude with a discussion of the current results from these searches and some predictions of how our
sensitivity will improve in the future.

\begin{figure*}
\centering\includegraphics[width=1.2\textwidth]{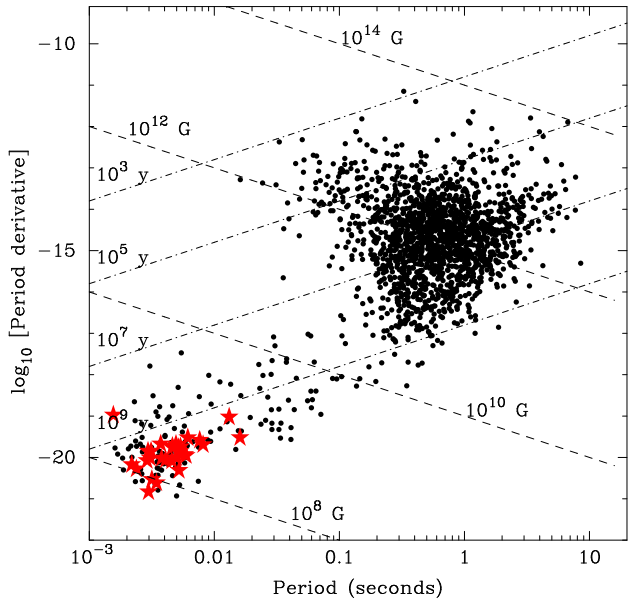}
\vspace{-0.5in}
\caption{Period vs period derivative diagram for pulsars. The MSPs that are currently included in the IPTA timing program
are marked as red stars.
The diagram shows lines of constant inferred surface dipole magnetic field ($B_{12} = 3.2\times10^{19} \sqrt{P\dot{P}}$) and constant
characteristic age ($\tau = P/2\dot{P}$). The MSPs have much lower magnetic fields and much larger characteristic ages than their non-recycled counterparts.
}
\label{fig:ppdot}
\end{figure*}

\section{Millisecond Pulsars}
\label{sec:msps}

Since the discovery of the first pulsar in 1967 \cite{hewish68}, over 2300 of these objects have been discovered\footnote{
http://www.atnf.csiro.au/people/pulsar/psrcat/},
with spin 
 periods ranging from 1.4~milliseconds to 8.5~seconds (see Figure~\ref{fig:ppdot}).
 Nearly 300 of these pulsars  have periods less than 20~ms.
These `millisecond pulsars', or MSPs\footnote{There is no strict cutoff for MSPs, with the main criterion being that they are fully recycled. However, there are no MSPs  with periods more than 20 ms included in PTAs.},  have been `recycled', or
spun-up to very short periods, through mass and angular momentum transfer from their binary companions. These pulsars
have spin-down rates orders of magnitudes smaller than their older, non-recycled counterparts (see Figure~\ref{fig:ppdot}). Their 
  very short periods allow us to measure the arrival times of their pulses to high precision, and their superb timing stability allows us to predict the pulse arrival
times with high accuracy.
Of the nearly 300 MSPs known, roughly 100 reside in globular clusters. They are not useful for GW detection experiments
due to the unknown, and not well-modeled, accelerations in the cluster. Therefore, roughly 200 Galactic MSPs are  possible additions to
PTAs. Recent MSP population studies show that there are a large number (roughly 30,000--80,000) of Galactic MSPs which are detectable, given sufficient sensitivity \cite{lor13,lbb+13}.
However, only a small fraction of these may have the brightness and timing stability necessary to be used in PTAs.

\begin{figure*}
\centering\includegraphics[width=0.7\textwidth]{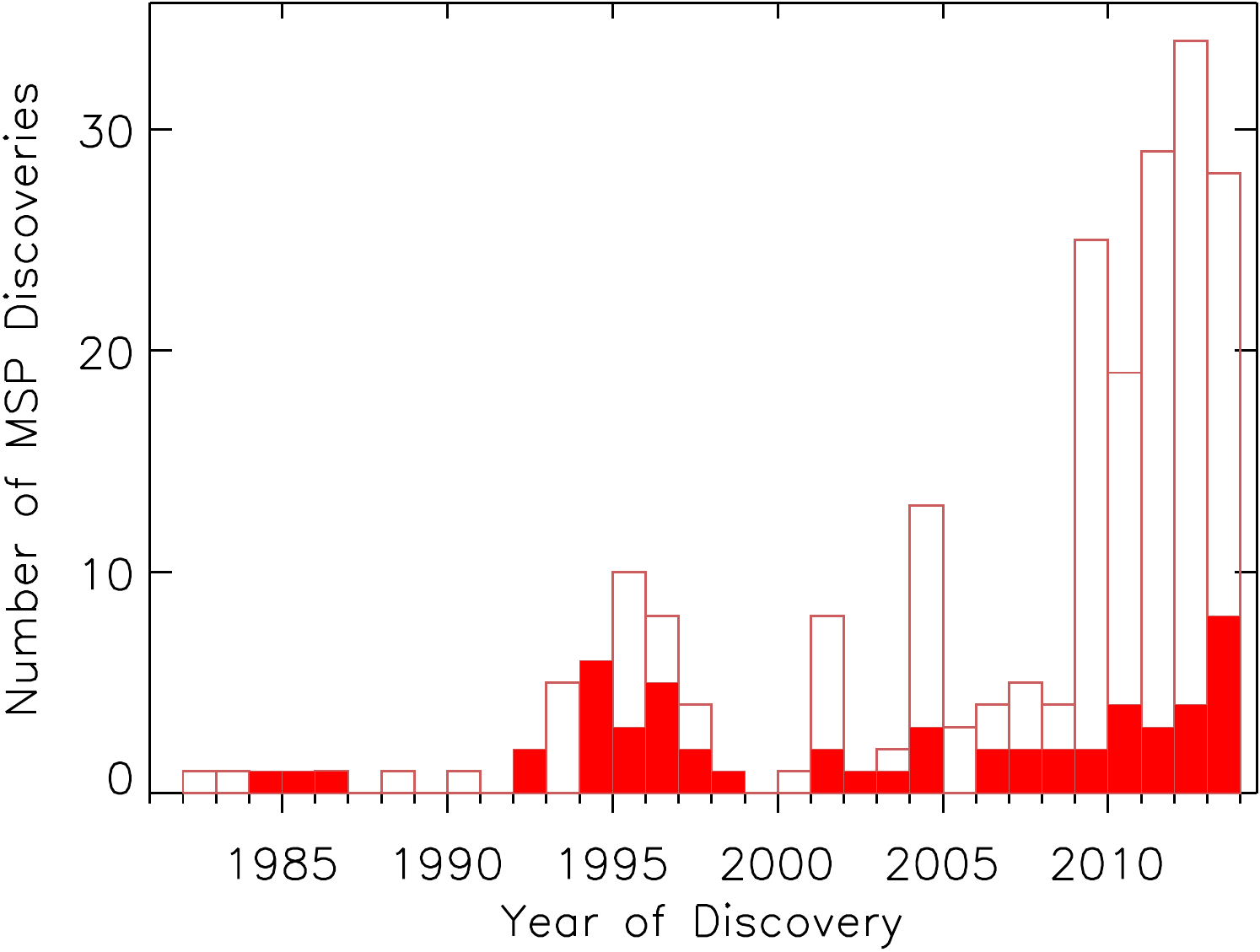}
\caption{
  The number of Galactic MSPs discovered vs year. The population has grown dramatically over the past several years. The red bars indicate the discovery years of the MSPs that are currently timed by the IPTA.}
\label{fig:nmsps}
\end{figure*}

In Figure~\ref{fig:nmsps}  we  illustrate the growth in the number of Galactic MSPs with time. The Galactic MSP population has {\it more than doubled} over the past several years due to very successful blind radio surveys and radio searches targeted at {\it Fermi} unidentified point sources. On average, 20--30 MSPs have been discovered yearly  since 2009 through a large international pulsar search effort in North America (with the Green Bank Telescope and the Arecibo Observatory; e.g. \cite{csl+12,lbr+13}), Europe (with the Effelsberg Telescope and the Nan\c{c}ay Radio Telescope; e.g. \cite{bgc+13,gfc+12}), Australia (with the Parkes Telescope; e.g. \cite{bbb+13}), and India (with the Giant Meterwave Radio Telescope; e.g. \cite{brr+13}). This trend is expected to continue, accelerating as MSP discoveries are made with the European Low Frequency Array \cite{sha+11} and, in several years, as the Chinese and South African FAST \cite{nlj+11} and MeerKAT telescopes \cite{bj12} are commissioned.

\begin{figure}
\centering\includegraphics[width=0.75\textwidth]{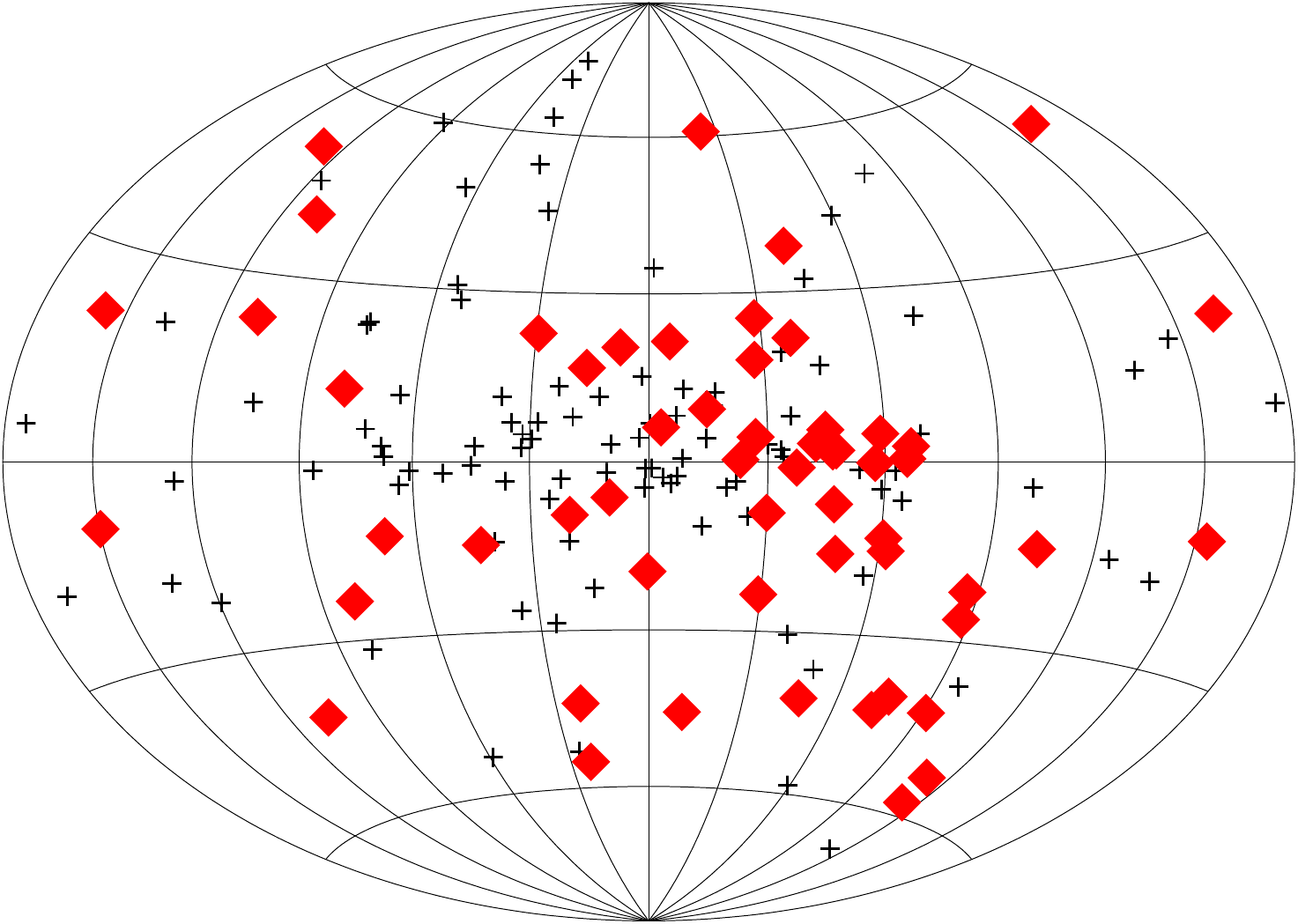}
\caption{Aitoff projection of Galactic MSPs (crosses), with those currently being timed by the IPTA marked as red diamonds.}
\label{fig:aitoff}
\end{figure}

In Figure~\ref{fig:aitoff} we illustrate the spatial distribution of the currently known MSPs; most of these have distances of order kpc (or thousands of light years); we are therefore largely sampling the local population with current surveys. Due to their large ages and relatively nearby distributions,  their spatial distribution is roughly isotropic, making them an attractive population for spatial correlation analyses such as
the gravitational wave detection experiment.

\vspace{-0.5in}

\newpage
\section{Calculating Timing Residuals}
\label{sec:timing}

\begin{figure*}
\centering\includegraphics[width=1.0\textwidth]{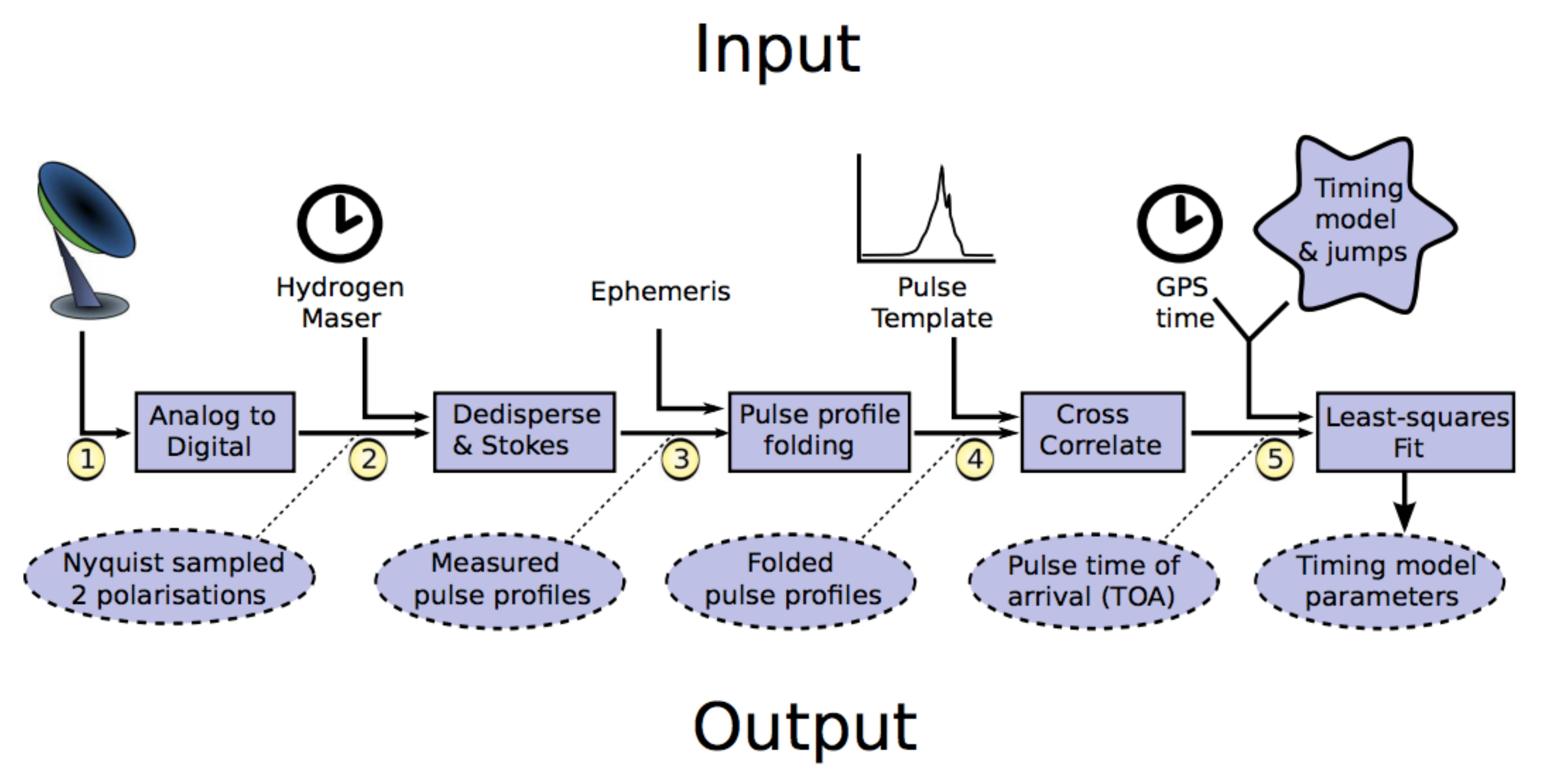}
\caption{
  The stages of pulsar timing, beginning with the collection of data at the radio telescope and culminating with the calculation of a timing model \cite{vlj+11}. In this Figure, adding the Stokes parameters comes before folding, but typically it is done after. Reprinted with permission of R. van Haasteren.}
\label{fig:timingsteps}
\end{figure*}

\begin{figure}
\centering\includegraphics[width=0.75\textwidth]{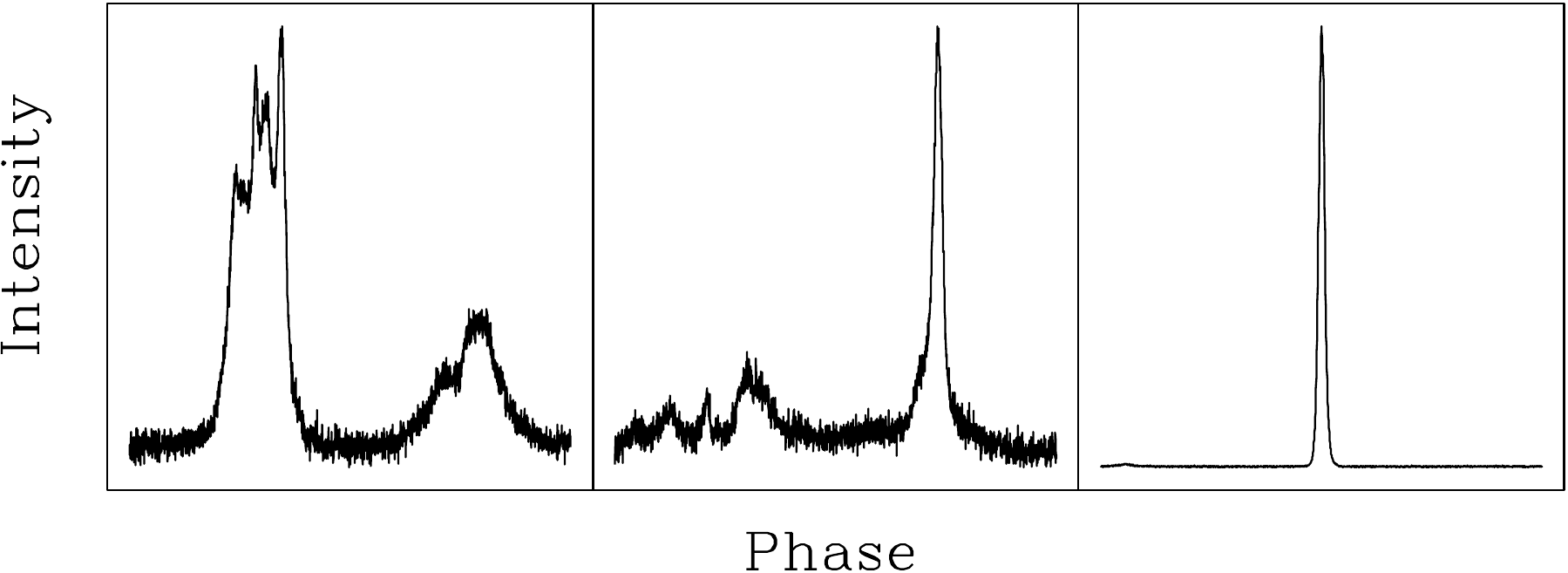}
\caption{Pulse profiles from one half-hour observation with the GBT at a center frequency of 1.4~GHz for three MSPs: (from left to right) PSRs~J0030+0451, J1614$-$2230, and J1909$-$3744. These show one full period for each pulsar. Credit: NANOGrav Timing Group.}
\label{fig:profiles}
\end{figure}

The first step in our GW detection experiment is to measure the times of arrival (or TOAs) of radio pulses from a collection of MSPs
as precisely as possible and to fit an accurate `timing model' to those TOAs. Figure~\ref{fig:timingsteps} illustrates the steps involved in this process;  a more
complete description may be found in Lorimer \& Kramer (2005) \cite{lk05}.
First, a pulsar is observed with a radio telescope and signals in two polarizations are digitized and recorded. The data are then dedispersed
to correct for frequency-dependent interstellar delays due to cold plasma dispersion by ionized free electrons. The delays are proportional to DM and $1/f^2$, where the dispersion measure
(DM) is the integrated column density of free electrons along the line of sight to the pulsar and $f$ is the center
frequency of observation.  The correction is typically done coherently
 by convolving the raw signal voltage with the
inverse of the  transfer function of the interstellar medium (ISM).   The data are then folded   with an  ephemeris containing an MSP's most up-to-date
 timing model to create full-Stokes profiles across the observation (i.e. one profile every 10~seconds to several minutes) and in a number of frequency subbands (i.e. typically 10 to 500, depending on the width of the radio bandpass). 
The  Stokes parameters are added to create  total intensity profiles (see Figure~\ref{fig:profiles}), using calibration observations to ensure that
the parameters are combined correctly.

A TOA is measured for each profile by cross-correlating a composite profile (i.e. summed over thousands to millions of individual pulses) with a high signal-to-noise template. This  template is typically the sum of pulse profiles over many epochs, from which the noise has been removed,
or a superposition of Gaussians  fit to the actual profile shape. 
These measured TOAs are then fit to a timing model through a least-squares fit \cite{ehm06}.  Traditionally, DM is fit using TOAs at multiple frequencies along with other parameters. However, new  methods that  measure a DM and TOA simultaneously demonstrate increased timing precision, in particular when wide bandwidths, over which pulse profiles evolve significantly, are used \cite{pdr14,ldc+14}. If multiple telescopes or observing backends are jointly fit it may be necessary to incorporate jumps, or phase offsets between sets of TOAs. Simultaneous observations with multiple telescopes should result in solving for these parameters and making their inclusion in timing models unnecessary \cite{dlc+14}.

\begin{figure}
\centering\includegraphics[width=1.00\textwidth]{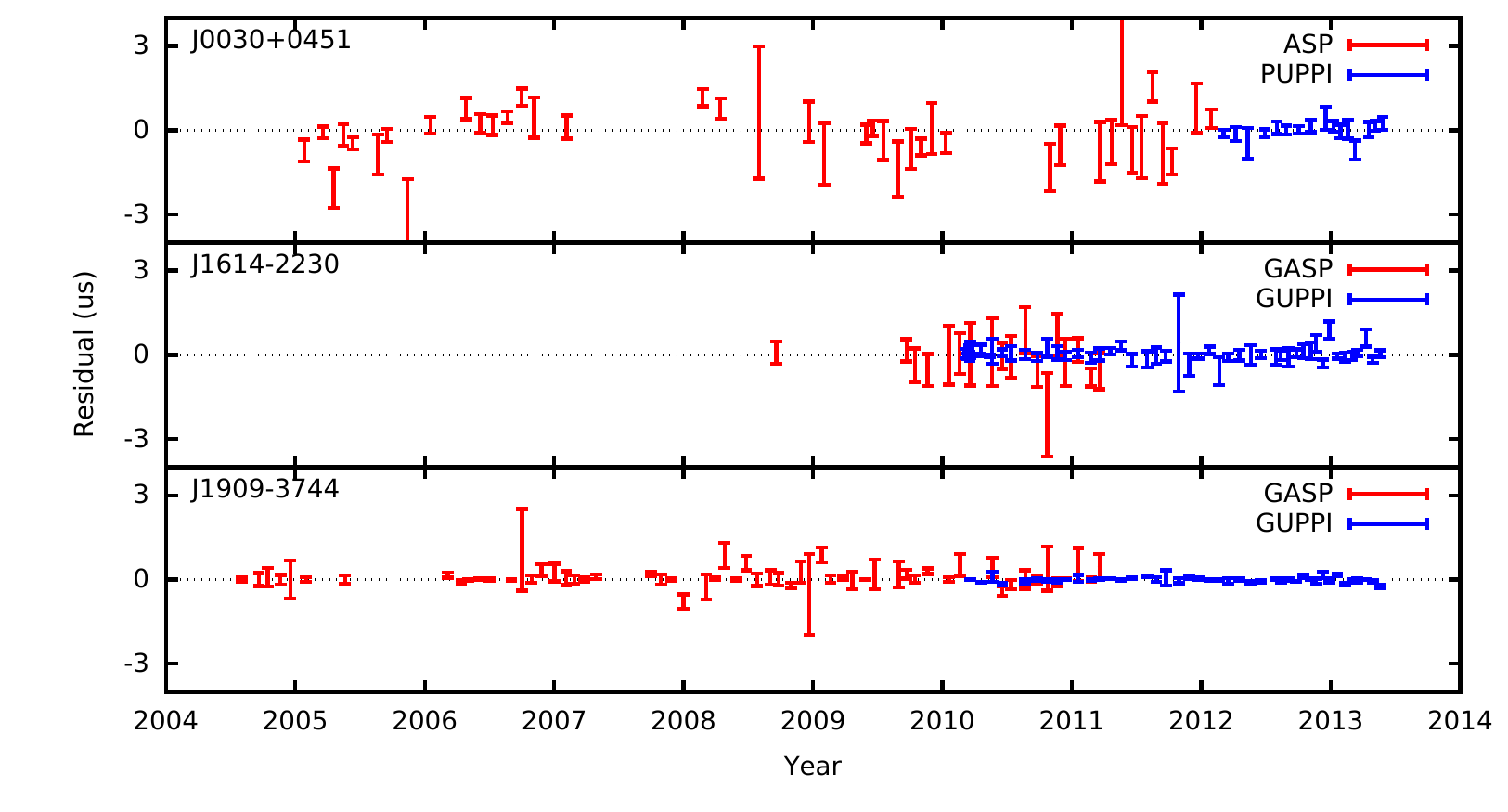}
\caption{Residuals for the same three MSPs shown in Figure~\ref{fig:profiles} timed by NANOGrav at a radio frequency of 1.4~GHz.
  Data taken with the  ASP/GASP backends, with 64-MHz bandwidth, are shown along with those taken using new backends GUPPI and PUPPI,  with 600--800 MHz bandwidths. The increase in bandwidth provides
 factors of two--three increased timing precision. 
Credit: Paul Demorest.
}
\label{fig:residuals}
\end{figure}

With each successive observation and fit, the timing model parameters will be measured more accurately and all known
effects will be accounted for. As time goes on, additional parameters will be required to account for lower
significance effects.
Some of the known effects are {\it extrinsic} to the pulsar;
they include the motion of the telescope within the
solar-system, including relativistic effects due to passage through the gravitational potential of the
Sun or other planets, and time variable dispersion. Measurement of the dispersion measure at each epoch
 requires TOA measurements with several days at multiple
radio frequencies, which are then fit  for DM. 
The timing model will also incorporate {\it intrinsic} effects such as the MSP's spin-down rate, position, 
motion through the sky (so-called proper motion), and parallax, as well
as orbital motion if the MSP is in a binary (as is the case for most MSPs). For some binary MSPs, relativistic effects
such as Shapiro delay or the orbital period derivative due to GW emission can be measured.
Once a timing
model for the TOAs of an MSP is achieved, ``timing residuals'' (see Figure~\ref{fig:residuals}) are
calculated by subtracting the model-predicted TOAs from the measured
TOAs. In addition to timing residuals which can be searched for the presence of GWs, the timing-derived parameters can be used
for a host of other experiments such as 
testing general relativity \cite{fw10}, measuring neutron star masses \cite{dpr+10,afw+13}, constraining binary evolution scenarios \cite{gsf+11},
and informing models of the interstellar medium \cite{kcs+13}.

There are many factors which determine the achievable timing precision. To first order, the precision will be
proportional to the ratio of the profile's width to its signal-to-noise.
Therefore, MSPs with narrow pulse profiles (e.g. Figure ~\ref{fig:profiles}) are most useful for this experiment.
The signal-to-noise of the profile will obviously be proportional to the brightness of the MSP at the frequency of observation.
 It will also depend  on the observing system's sensitivity,
 which scales linearly with collecting area and with the square roots of integration time and
bandwidth \cite{lk05}. {\it Therefore, TOA precisions are greater for bright MSPs with narrow pulses observed with large telescopes for long integration times  and wider bandwidths.
}

In addition, there are several more subtle considerations.
 The timing precisions of a large number of pulsars being timed by Parkes, for instance, are limited by instrumental polarization artifacts,
making proper polarization calibration critical \cite{vanstraten13}. In addition, there are multiple methods for correction of frequency dependent effects, such as dispersion and
scattering. Dispersion effects must be removed with care, as simple methods for time-variable DM fitting could remove astrophysics signals of interest \cite{kcs+13}.
Interstellar scattering due to inhomogeneities in the interstellar medium can also cause frequency dependent delays which  may vary by an order of magnitude more than the timing precisions required for GW detection \cite{hs08}.
These delays can be accurately estimated through sophisticated deconvolution techniques such as cyclic spectroscopy \cite{demorest11}, but the range of applicability of these techniques is not yet clear.
Furthermore, some MSPs show evidence for 
intrinsic spin noise. This noise has a very `red' power spectrum, where most of the power is concentrated at low frequencies.
Red spin noise is well studied in slower, non-recycled pulsars, and power spectral indices (i.e. $P(f) = f^{\alpha}$) of $\alpha = -3$ to $-5$ have been  measured \cite{ddm97}. There is little evidence for substantial amounts of spin noise for MSPs \cite{vbc+09,dfg+13}, but it
could begin to play a more limiting role as the lengths of the  observing campaigns increase. While there is evidence that spin noise is greater for non-recycled pulsars
with higher spin-down energy loss ($\dot{E}$) rates \cite{sc10}, it is not yet clear whether that relationship holds for MSPs, and it is unlikely that it would be sufficiently predictive to exclude high $\dot{E}$ pulsars from PTAs. 

\section{Searching for and Characterizing Gravitational Wave Sources}
\label{sec:searching}

Gravitational waves cause the light travel times between a pulsar and the Earth to vary. At a particular time,
the total time delay induced by a GW source will depend on the GW strain at the pulsar at the time of the radio wave emission (i.e. the `pulsar term'), the GW
strain at the Earth at the time of the radio wave reception (i.e. the `Earth term'), the orientation of the source, pulsar, and Earth, and the GW frequency, with
the induced residual inversely proportional. With this scaling, and an average timing precision of 100~ns over a
total timespan of 5--10 years, a PTA would be sensitive to strains of order $10^{-15}$  to $10^{-16}$ \cite{jhl+05}.
This effect will be detected as a characteristic directional signature in the timing residuals with a
  quadrupolar angular
correlation on the sky \cite{hd83}, with the exact 
 signature  varying depending on the type of GW source. 
For all cases,  GW searches will be performed on  timing residuals (calculated as described in Section~\ref{sec:timing}) or
by
maximizing~\cite{esvH13} or marginalizing~\cite{vlm+09} the likelihood
over the timing model parameters in such a way that the timing model
parameter measurements and GW wave searches are performed
simultaneously.
 
\begin{figure}
\centering \includegraphics[width=0.85\textwidth]{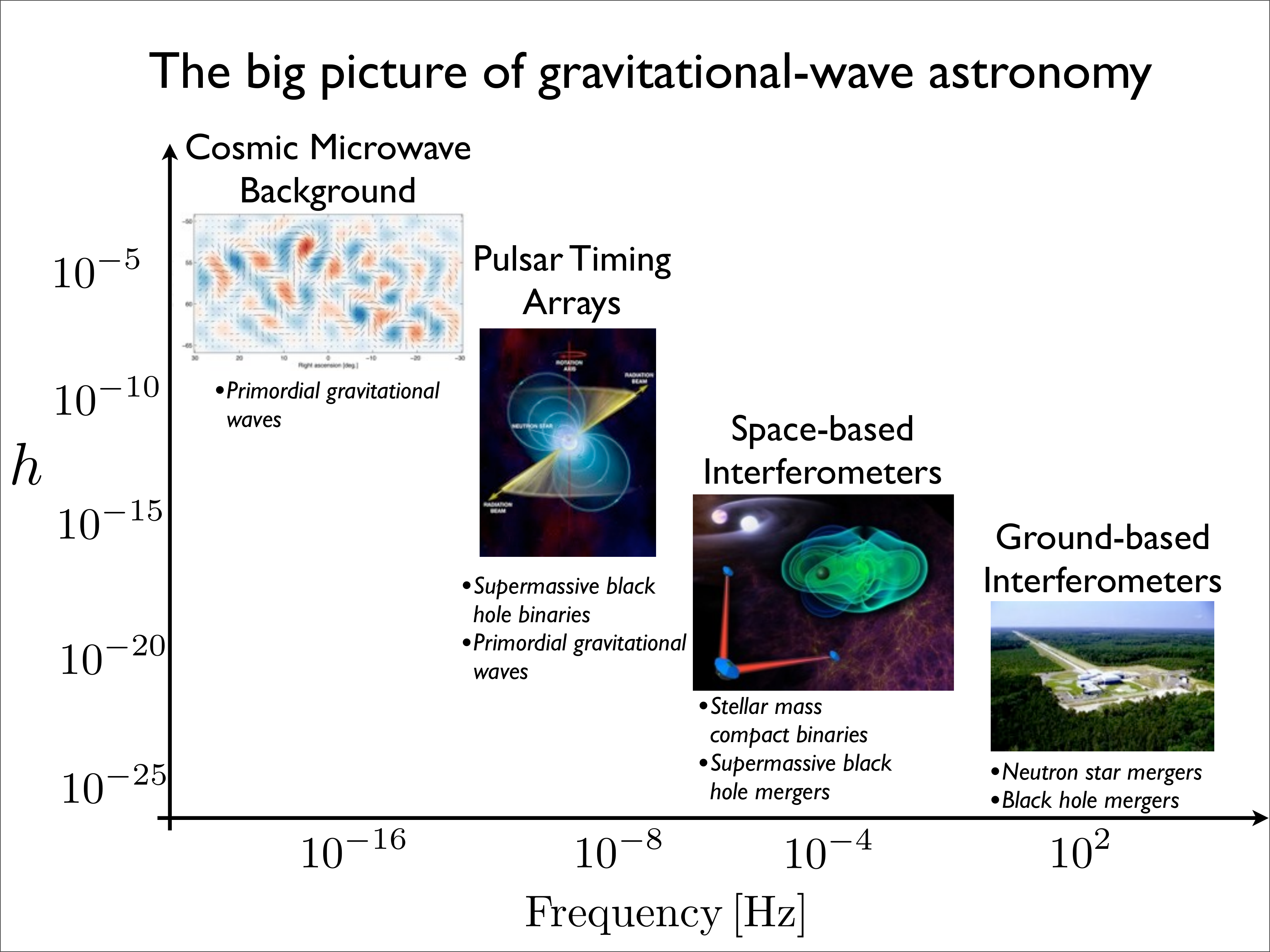}
\caption{The range of amplitudes, measured by strain $h$, and  frequencies to
which the four essential GW detection techniques are sensitive are shown, along with the
 expected  sources that would produce GWs at these amplitudes and frequencies.
 Cosmic microwave background polarization experiments will aim to confirm the  B-mode polarization detection. At higher GW frequencies, experiments are targeting astrophysical sources. PTA experiments include NANOGrav, the EPTA, and the PPTA, which together form the IPTA.
 A space-based interferometer called eLISA (Evolved Laser Interferometer Space Antenna) is currently in the development phase, with
a planned European Space Agency launch date of 2032. Ground-based GW interferometers include LIGO,
(USA),  VIRGO (Italy/France), TAMA300 and LCGT (Japan), and GEO600 (Germany/U.K.).  The most sensitive of these experiments, LIGO,
which consists of two detectors, one in  Louisiana
and one in Washington, is currently undergoing a sensitivity upgrade to become Advanced LIGO. Credit: NANOGrav}
\label{fig:bigpicture}
\end{figure}

The sensitivity of a pulsar timing array peaks at frequencies corresponding to the inverse of
the total timespan of the data. For example, the minimum detectable strain for an
experiment with a duration of 10 years will be at a frequency of $3\times10^{-9}$~Hz.
At frequencies higher than this, the signatures induced in pulsar timing residuals are smaller because of the intrinsic inverse dependence on GW frequency \cite{detweiler79}. At frequencies lower than this, the signal would be largely absorbed in timing model fits.
In Figure~\ref{fig:bigpicture}, we illustrate how the frequencies at which PTAs are sensitive compare to those of other GW detection experiments.

\begin{figure}
\centering\includegraphics[width=0.85\textwidth]{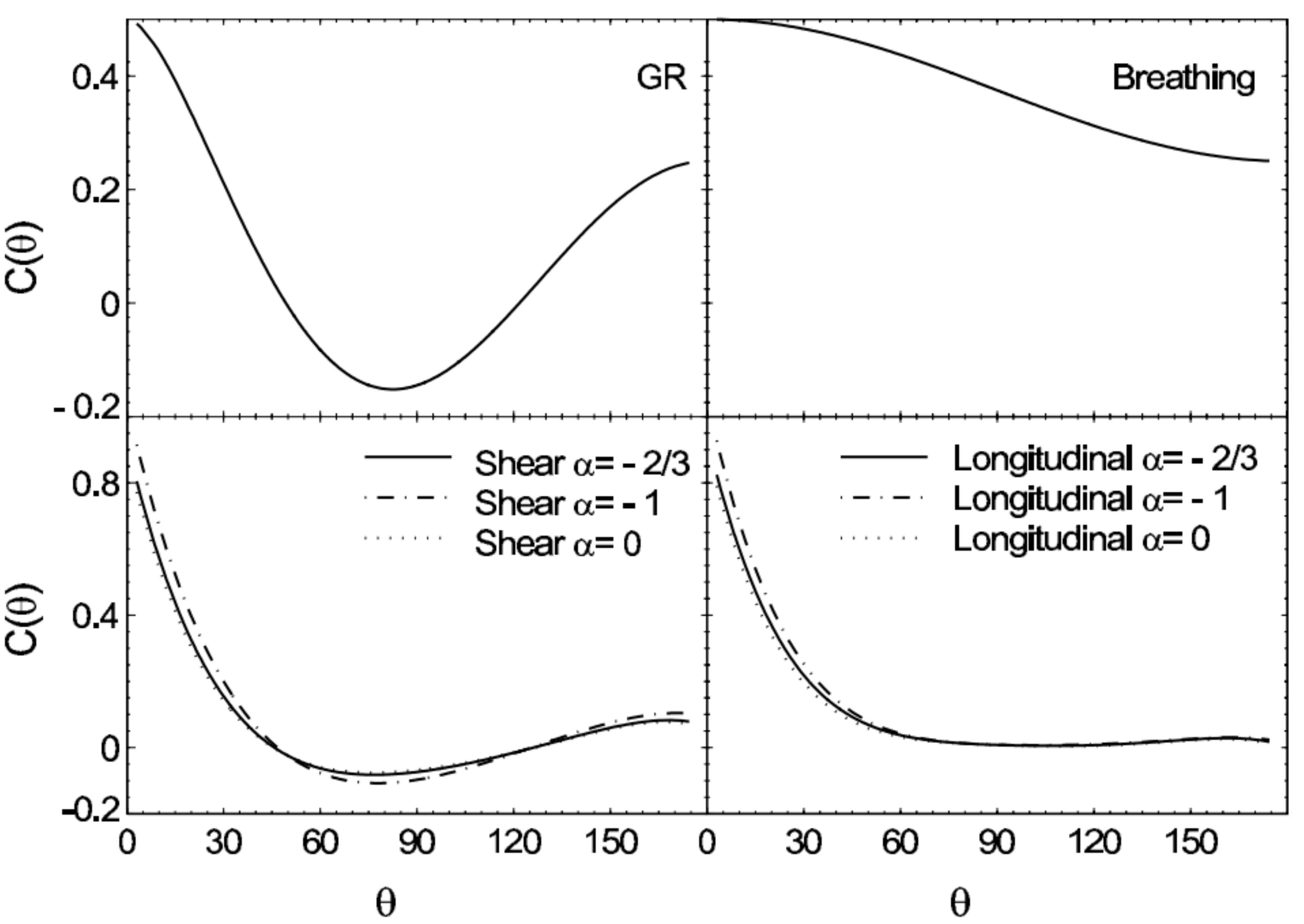}
\caption{The expected correlation of pulse residuals vs angular separation for isotropic, stochastic distributions of
GW sources under different assumptions about gravity. The top left figure, known as the `Hellings and Downs curve', assumes that general relativity is correct \cite{hd83}. The other three figures show the expected correlation in the cases that GWs have
breathing, shear, or longitudinal modes not predicted by general relativity, where $\alpha$ is the spectral index of the GW background \cite{ljp08}. For SMBH binaries in the frequency range of interest to PTA, $\alpha$ is expected, under the most simplistic assumptions, to be $-2/3$ \cite{jb03}. Reprinted with permission of K. J. Lee.
}
\label{fig:hd}
\end{figure}

PTAs are sensitive to stochastic, continuous, and burst GW sources. Stochastic
sources are not resolvable as discrete signatures and can either be due to the superposition of many individual sources
or due to a true background  of GWs.
Figure~\ref{fig:hd} illustrates the expected correlation in residuals vs. angular separation \cite{hd83} for an
isotropic distribution of GWs.
In this case, only the Earth terms are correlated, with a maximum correlation coefficient of 0.5.
The stochastic background detectable by PTAs likely includes contributions from SMBH binaries \cite{sesana13} and, possibly, cosmic strings \cite{Siemens:2006vk},  early Universe phase transitions~\cite{ccd+10}, and relic GWs from
inflation~\cite{PhysRevD.37.2078,tong14}. {\it All of these would provide 
unique windows into cosmology and galaxy formation and evolution \cite{sesana13_2}.
}

A
stochastic background of any kind will manifest itself with a characteristic strain spectrum like $h_c(f)=A(f/{\rm yr}^{-1})^\alpha$,
where $A$ is the amplitude of the GW wave background and $f$ is the GW frequency. Assuming that the binaries are circular and that they are losing
energy and angular momentum only to GWs, it is straightforward to show that $\alpha = -2/3$ \cite{rr95,phinney01,jb03};
 this predicted slope is independent of the cosmology assumed. While shorter orbital period objects have larger GW amplitudes, they also have much shorter lifetimes, meaning that most of the GW power is concentrated at low GW frequencies.
 The power spectrum of this background will then equal $P(f) = h_c(f)^2/(12\pi^2 f^3)$, i.e. scaling  as $f^{-13/3}$ for a SMBH binary background \cite{jhs+06}.  It is sometimes useful to discuss stochastic background
strengths in terms of their fractional contribution to the energy density of the Universe per logarithmic frequency interval,
 which is defined as $\Omega_{\rm GW}(f) = 2\pi^2 f^2 h_c^2/(3 H_0^2)$, where $H_0$ is the Hubble constant \cite{jhs+06}.
 
Other possible sources of stochastic backgrounds are also expected to have `red' (i.e. with more power at low frequencies)
spectra, but with different predicted $\alpha$ values. For a background due to interacting cusps and loops on cosmic strings, $\alpha$ is expected to be roughly $-7/6$ \cite{dv05}. Relic radiation from the early universe is predicted to have a $\alpha \simeq -1$ \cite{grishchuk05}. 
For any expected scenario, therefore,
we expect stochastic GWs to manifest themselves in pulsar timing residuals
as red noise. 
 Upper limits on the stochastic GW background can be
calculated based solely on the amount of red noise seen in power spectra of the timing residuals of individual pulsars \cite{ktr94,src+13}.
Some upper limits, and necessarily any detections, however, rely
 on calculating the correlation between the residuals of pairs of pulsars as a function of angular
separation and placing limits on the strength of 
 the expected quadrupolar signature \cite{dfg+13}.
These upper limits assume that general relativity is correct and that the expected correlation is that shown in the top left panel of Figure~\ref{fig:hd}. 

Note that the particular shape depicted in Figure~\ref{fig:hd} is only expected for the case of an {\it isotropic} distribution of GW sources, assuming that general relativity is correct. Anisotropic backgrounds or individual sources could also be detected through correlation analyses which incorporate a range of correlation function shapes \cite{cn13,tg13}.
Similarly, non-Einsteinian longitudinal, breathing, or shear polarization models would manifest themselves
as deviations from the Hellings \& Downs  curve in Figure~\ref{fig:hd}.
Furthermore, the slope of the stochastic background could  differ from the canonical $-2/3$ at the frequencies of interest if binaries are eccentric or if
stellar interactions are taken into account \cite{rws+14}. 

\begin{figure}
\centering\includegraphics[width=0.85\textwidth]{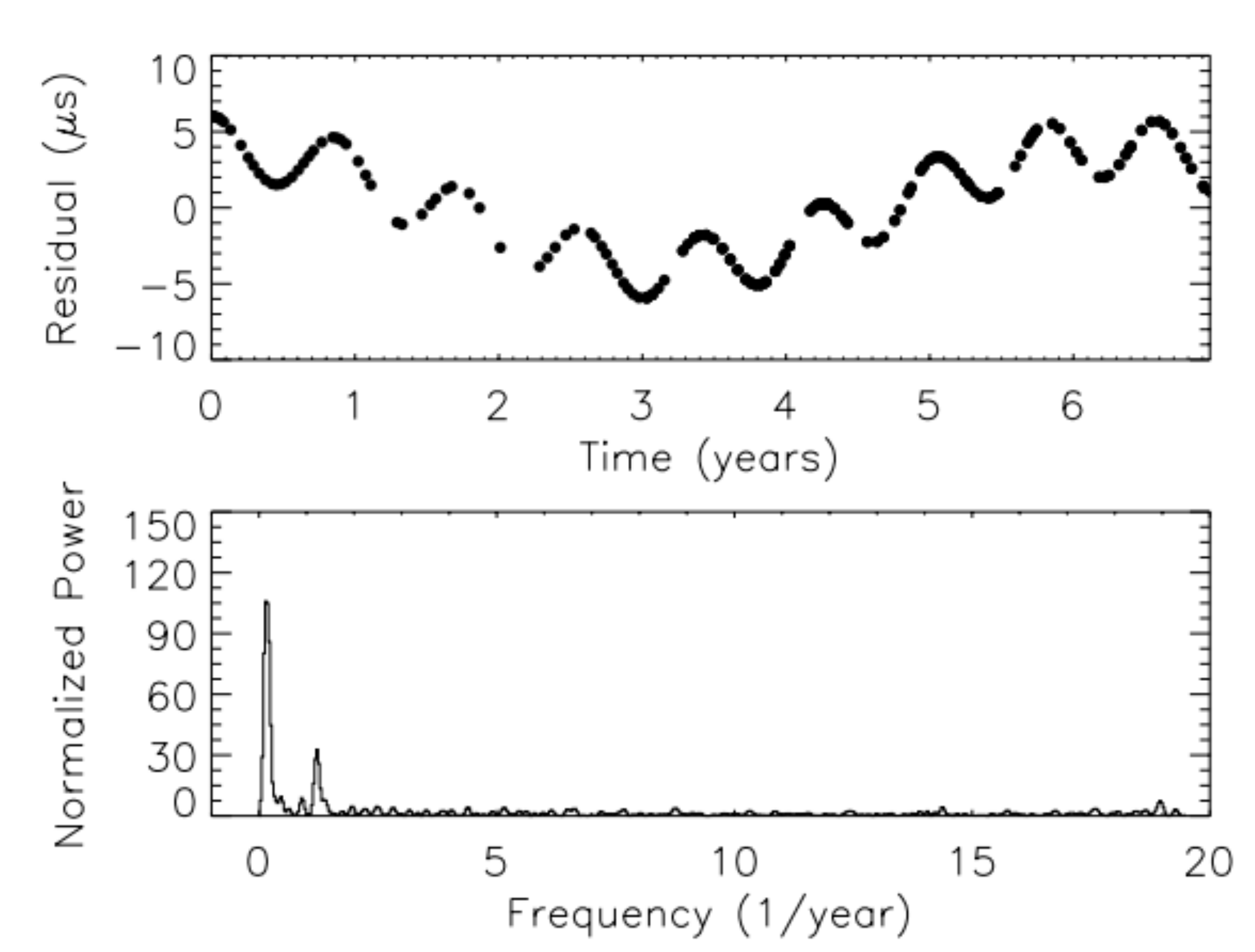}
\caption{Top: The expected residuals induced in MSP B1855+09 due to a proposed supermassive black hole binary \cite{simt03} with combined (upper limit) mass of 50~billion~M$_\odot$ and orbital period of one year in the galaxy 3C66B \cite{jllw04}.  Bottom: Power spectrum of the residuals showing the clear Earth and pulsar term modulation. More recent work suggests that the mass of this binary system is much less than the upper limits originally presented by a factor of roughly 50 \cite{ios10}. Reprinted with permission of R. Jenet.}
\label{fig:continuous}
\end{figure}

Continuous GW sources emit at a single (but evolving) GW frequency.
Supermassive black hole binaries are the most likely source of continuous GWs for PTAs. Due to loss of energy
from GW emission, we expect the
GW frequency to evolve with time so that the frequency of the Earth term will be higher than the frequency of the pulsar term (see Figure~\ref{fig:continuous}).
 However, in order to include the pulsar term in a continuous wave search, the pulsar distance must be known to within a fraction of a
GW wavelength. This is not possible for any MSPs thus far (though one is within a factor of two of this goal \cite{dbl+13}); however, while computationally expensive,
 the pulsar distance can be included in continuous wave searches as a search parameter, simultaneously allowing improved GW source localization and, possibly, better pulsar distance measurements \cite{lwk+11}. 
 Even without the inclusion of the pulsar distances as a search parameters, continuous wave searches can be extremely computationally intensive due to the large range
of possible binary parameters.
Bayesian and frequentist techniques that can determine the amplitude, period, and position (albeit to limited precision) of a candidate GW source are currently
being explored by various groups \cite{bs12,esc12}.

\begin{figure}
\centering\includegraphics[width=1.0\textwidth]{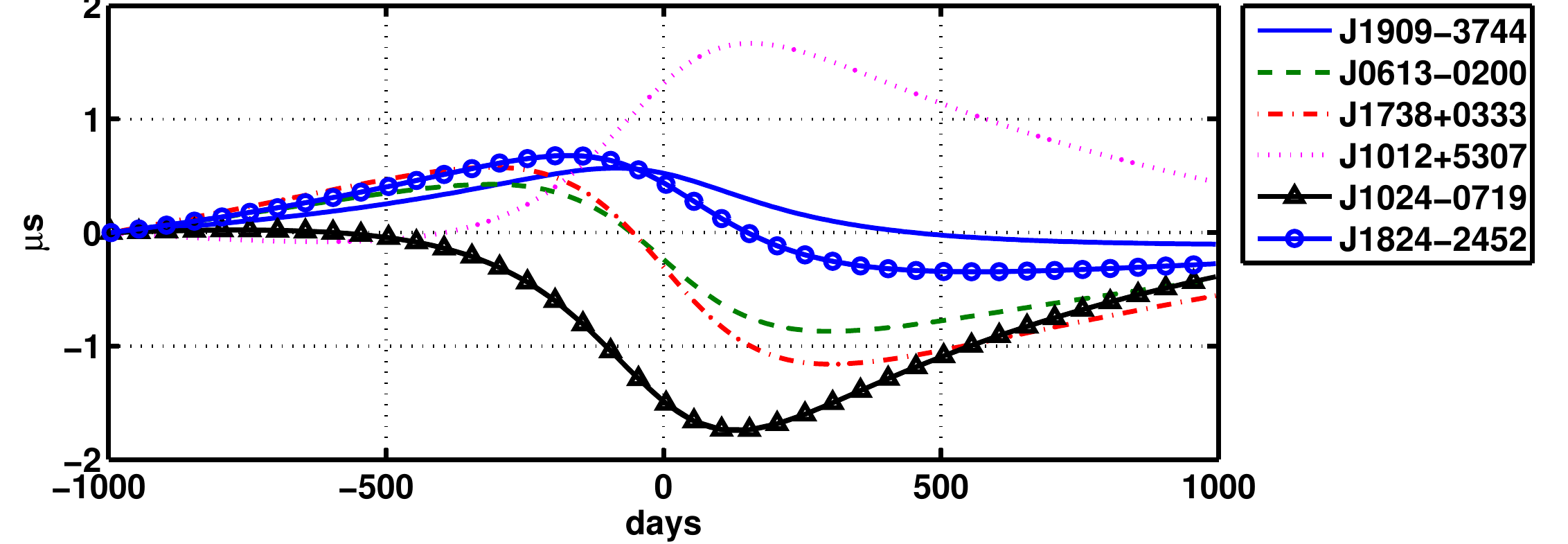}
\caption{The timing residuals induced in a sample of IPTA pulsars for the
 parabolic encounter of two 10$^{9}$~M$_{\odot}$
 black holes, with an impact parameter of 0.02 pc, at a distance of 15~Mpc in the direction of the Virgo
Cluster of galaxies \cite{fl10}. Reprinted with permission of L. S. Finn.}
\label{fig:burst}
\end{figure}

Burst sources have signal durations much shorter than the total time spans of observations
and could be due to mergers of SMBHs,
 periastron
passages of compact objects orbiting a SMBH, or cusps on cosmic strings \cite{dv00}.  Bayesian pipelines have been constructed to detect burst sources, even when
the waveform cannot be determined or the source localized \cite{vl10,fl10}; see Figure~\ref{fig:burst} for an example of an expected burst signature in timing residuals. Recent work shows that because burst signals grow with data span,
  red noise can hinder the detection of bursts and, likewise, bursts could make the stochastic GW background more difficult to detect \cite{cj12}.
Algorithms are also being developed to search for non-oscillatory, permanent deformations of space-time known as bursts with memory (BWMs). These would have the appearance of a `ramp' function in pulsar timing residuals that, like all GW signatures, would show a quadrupolar angular correlation and appear in the Earth term of
multiple pulsars \cite{mcc14}.
Developing accurate noise models for pulsar residuals will be critical to their detection \cite{mcc14}.

\section{Current Timing Programs}
\label{sec:observations}

There are currently three collaborations which are using timing observations of MSPs to search for and characterize GWs.
The Parkes Pulsar Timing Array, or PPTA, was formed in 2004 and undertakes observations with the 64-m Parkes Telescope in NSW, Australia. Currently, 20 pulsars are observed roughly every two--three weeks at radio frequencies of 0.7, 1.4, and 3~GHz.
All data are coherently dedispersed 
 over bandwidths of 64~MHz at 700~MHz, 300~MHz at 1.4~GHz, and 1~GHz at 3~GHz \cite{mhb+13}.

 The European Pulsar Timing Array, or EPTA, was formed in 2005 (though the collaboration existed before that time) and uses five telescopes in Europe. These include the 76-m Lovell Telescope in the UK, the 100-m Effelsberg Telescope in Germany, the 95-m (effective diameter) Nan\c{c}ay Radio Telescope (NRT) in France, the 93-m (effective diameter) Westerbork Synthesis Radio Telescope (WSRT) in the Netherlands, and the newly commissioned 64-m Sardinia Radio Telescope (SRT) in Italy. When all five telescopes operate together as the Large European Array for Pulsars (LEAP), they synthesize a 194-m equivalent dish to provide very high-precision data.
The EPTA collaboration observes 60 MSPs at roughly monthly cadences with all five telescopes \cite{kc13}. Note that since the majority of MSPs are observed with
multiple telescopes, the effective cadence is far greater than monthly.  The observations are primarily at 1.4~GHz,
with low-frequency (350-MHz) WSRT observations and $>$2~GHz Effelsberg, NRT, and WSRT observations supplementing these to facilitate DM correction. All observatories employ coherent dedispersion backends with bandwidths of 400--500~MHz at 1.4~GHz. 
An ultra broadband receiver, which will cover the entire frequency range from 0.6--3~GHz, is currently being
commissioned at Effelsberg. If this receiver performs as expected, if will not only increase the sensitivity, but also enable
more accurate DM correction.

The North American Nanohertz Observatory for Gravitational Waves \\
(NANOGrav) was formed in 2007 and uses the 300-m Arecibo Observatory in
Puerto Rico and the 100-m Green Bank Telescope (GBT) in West Virginia, USA. NANOGrav observes 43 pulsars at frequencies of 800 and 1.4~GHz with the GBT and at two of 0.4, 1.4, and 2~GHz at Arecibo \cite{mclaughlin13}. Data are coherently dedispersed over bandwidths of
50~MHz at 430 MHz, 200~MHz at 800~MHz, and 800~MHz at higher frequencies. 
Observations occur at each frequency and each telescope roughly every three weeks, with two pulsars observed at both telescopes.
 Due to its large size and hence unparalleled sensitivity, 
all of the MSPs that are within Arecibo's declination range (i.e. between $-1^\circ$ and $+38^\circ$) are
observed at Arecibo. The GBT, with its much greater sky coverage (i.e. all declinations north of $-45^{\circ}$)
 is used for the remainder of the pulsars.

\begin{figure*}
\centering\includegraphics[width=1.0\textwidth]{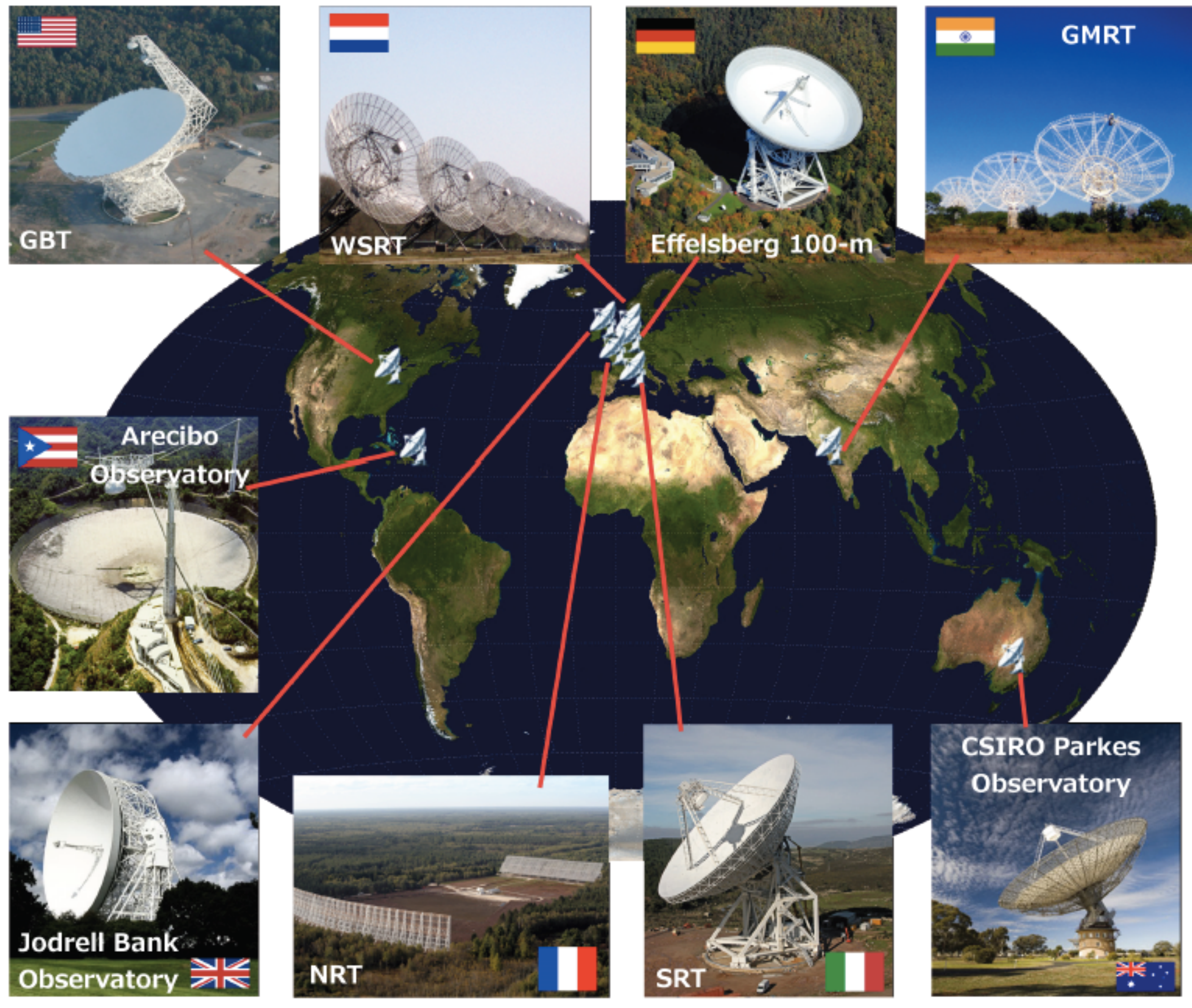}
\caption{The telescopes used for IPTA observations. Though not officially part of any PTA's standard timing program, the GMRT is included as it has begun a low-frequency DM monitoring campaign for IPTA pulsars.  Credit: Brian Burt.}
\label{fig:iptatelescopes}
\end{figure*}

\begin{figure*}
\centering\includegraphics[width=1.0\textwidth]{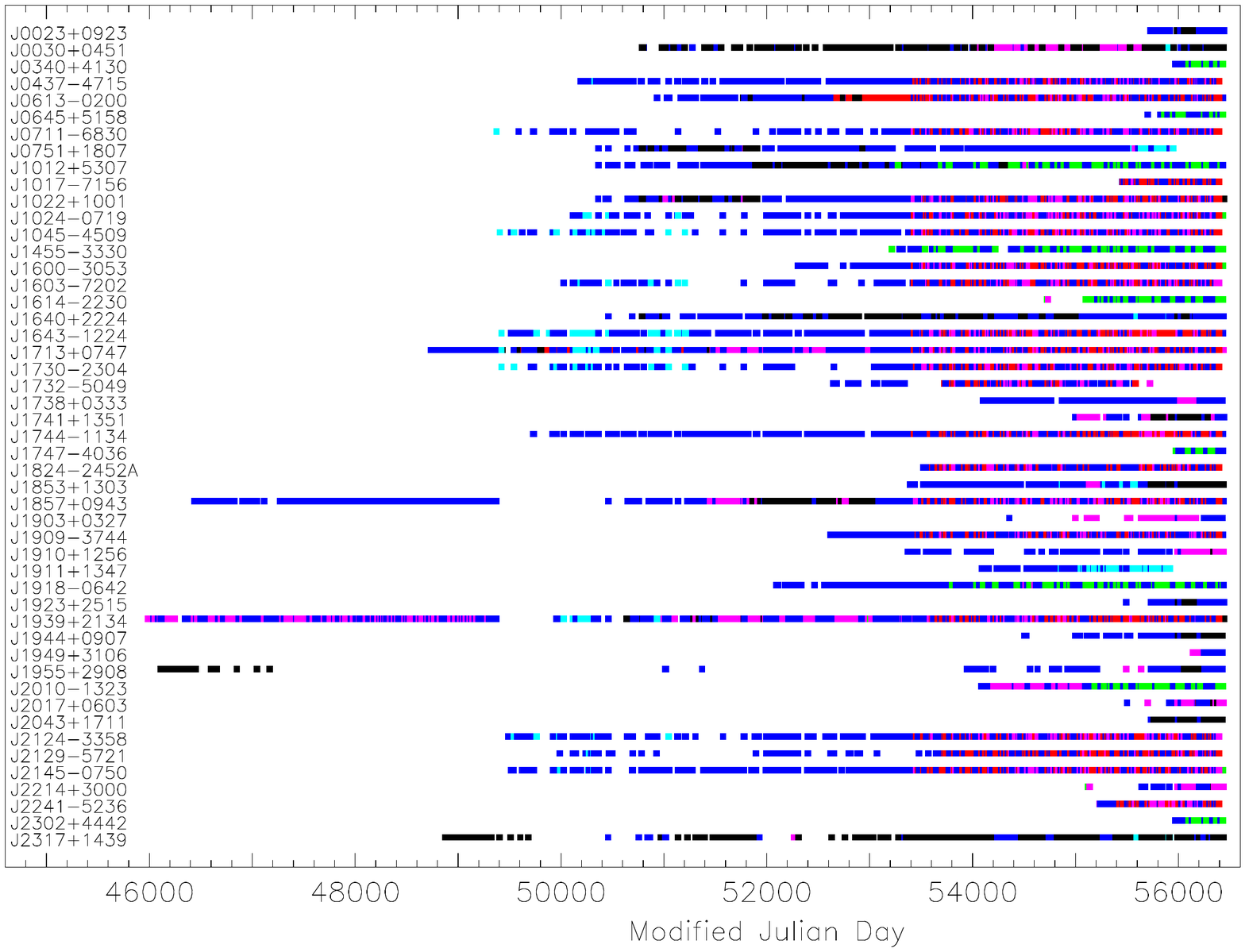}
\caption{Distribution of TOAs for the combined IPTA data release \cite{manchester13}. The center
frequencies of observation are color coded as follows: black: $<$ 500 MHz, red:  between 500 and 750 MHz, 
 green: between 750 and 1000 MHz, blue: between 1 and 1.5 GHz,
aqua: between 1.5 and 2 GHz, and pink: between 2  and 4 GHz. Reprinted with permission from Dick Manchester and IOP Publishing.   All rights reserved.} 
\label{fig:release}
\end{figure*}

The PPTA, EPTA, and NANOGrav are all part of the International Pulsar Timing Array (IPTA), which uses all of the most sensitive radio telescopes in the world to facilitate the goal of  GW detection  (see Figure~\ref{fig:iptatelescopes}).  The growth in the number of MSPs being timed through IPTA efforts, and their current sky distribution, are illustrated in Figures~\ref{fig:nmsps} and \ref{fig:aitoff}. While an IPTA data sharing agreement was signed in 2009,
combining data from all three PTAs and all seven (soon to be eight, with the SRT) telescopes is non-trivial due to clock
offsets and varying data formats, but the first IPTA data release will be published over the next several months. Figure~\ref{fig:release} illustrates the data that will initially be available to all IPTA members for GW analyses.
The IPTA ratified a publication policy in 2012 that outlines procedures for proposing projects involving the full IPTA data set. Several of these are already proposed and the first GW limits using IPTA data should be published within the year.

\section{Current Stochastic Background Limits}
\label{sec:sb}

\begin{figure*}
\centering\includegraphics[width=0.7\textwidth]{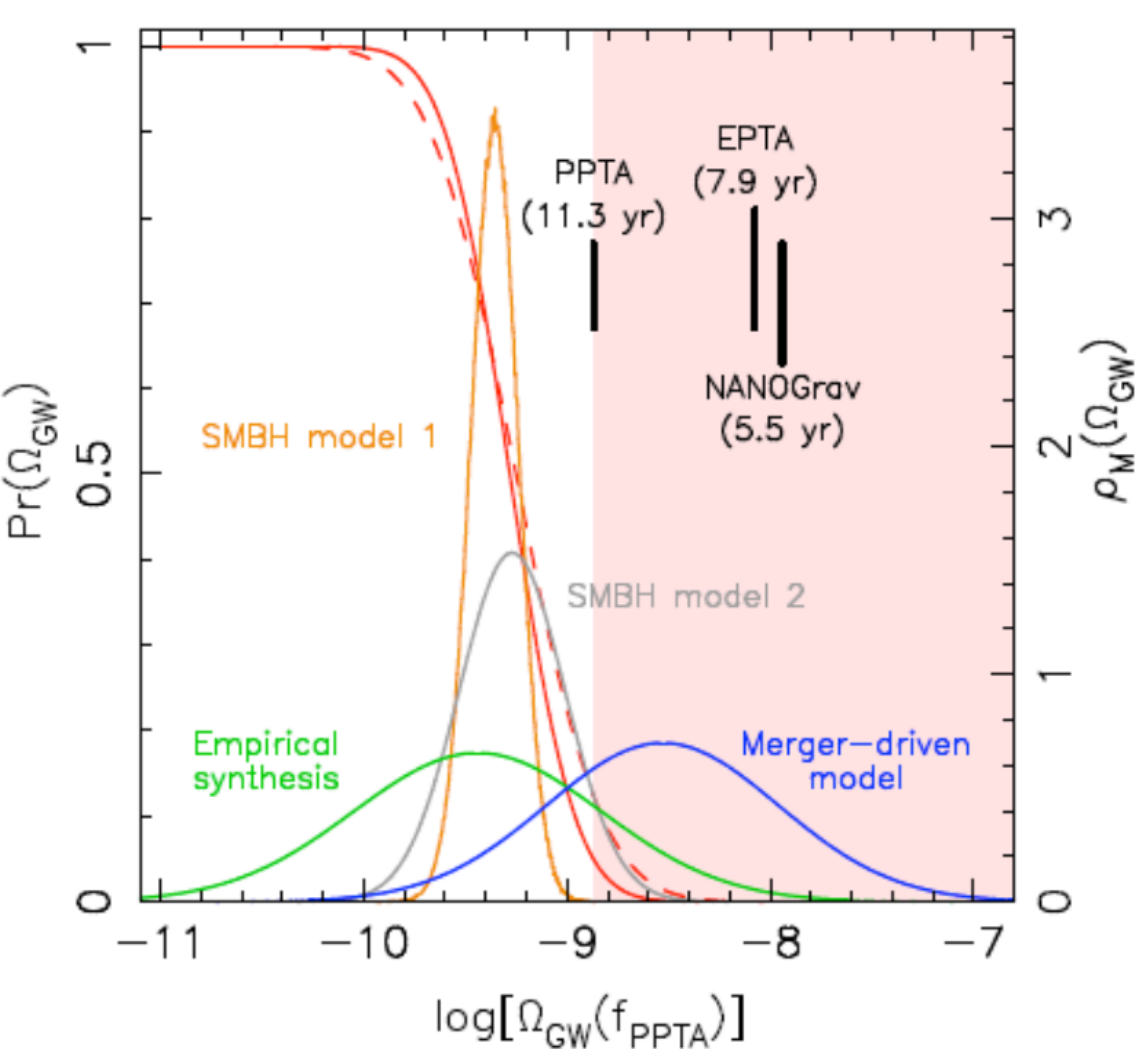}
\caption{A comparison between the various PTA constraints on $\Omega_{GW}$ and predictions from several models \cite{src+13}. The vertical bars illustrate the 95\% confidence upper limits on $\Omega_{GW}$ scaled to a GW frequency ($f_{\rm PPTA}$) of 2.8~nHz for the three most recently published PTA upper limits \cite{src+13,vlj+11,dfg+13}. These limits are based on datasets of length 11.3, 7.9, and 5.5 years. The pink shaded area represents values of $\Omega_{GW}$ that are excluded with 95\% confidence. The red curve illustrates the regime
ruled out by the recent PPTA limit, with the solid line assuming a Gaussian GW background and the dotted line a non-Gaussian background.
 The other four curves are the predictions for four other models. These include a purely merger-driven model \cite{mop14}, a model synthesizing 
 several current observational estimates \cite{sesana13}, a semi-analytic model (SMBH model 1) based on the the Millennium \cite{swj+05}  and Millennium II \cite{bsw+09} dark matter simulations \cite{src+13}, and a model (SMBH model 2) assuming that gas accretion is the dominant source of mass accumulation in nearly all  SMBHs  \cite{kon+13}. The $\Omega_{GW}$ limits correspond to $h$ (at a frequency of yr$^{-1}$)
 limits of $7\times10^{-15}$, $6\times10^{-15}$, and $2.4\times10^{-15}$ for NANOGrav, EPTA, and PPTA, respectively.  
Reprinted with permission of R. Shannon and AAAS. All rights reserved.}
\label{fig:limits}
\end{figure*}

All three PTA collaborations are performing several different kinds of GW analyses that are sensitive to stochastic, burst, and continuous wave sources. The most straightforward means to compare the three efforts, and gauge the overall sensitivity of the worldwide endeavor,  is to review the upper limits placed by the three groups on the GW stochastic background.
The 95\% confidence upper limits on the characteristic strain $h_c$ at a frequency of yr$^{-1}$ are
$2.4\times10^{-15}$, $6\times10^{-15}$, and $7\times10^{-15}$ for the PPTA, EPTA, and NANOGrav, respectively \cite{src+13,vlj+11,dfg+13}. These limits, illustrated in Figure~\ref{fig:limits}, are based on datasets of length 11.3, 7.9, and 5.5 years, respectively. All assume that the stochastic GW background has a spectral slope of $-2/3$.
 
The PPTA limit
was calculated from the power spectra of the six PPTA pulsars with the highest timing precisions. These power spectra were modeled
with a white noise component, a red noise component due to the GW background, assumed to be due to SMBH binaries, and (for one pulsar) an additional red noise component corresponding to intrinsic spin noise.
 While providing the most sensitive upper limit yet, this approach
 could not result in a detection as no spatial correlation analysis was performed due to the lack of a
sufficient number of high timing precision pulsars.

Both of the
  EPTA and NANOGrav limits were based on analyses that constrain the presence of the expected angular correlation due
to a GW background (as in Figure~\ref{fig:hd}). The EPTA limit was calculated through a Bayesian analyses which marginalized over the timing model parameters, white and red noise terms intrinsic to the pulsars and the various telescopes, GW background amplitude, and GW background spectral index.  This analysis was performed using data from  five pulsars observed with three radio telescopes. While the quoted 95\% confidence upper limit of $6\times10^{-15}$ assumes $\alpha = -2/3$, the spectral slope was a free parameter in the analysis. 

The NANOGrav limit was calculated in a different way, with the timing model fit performed first and then post-fit residuals analyzed for the presence of the expected angular correlation, using the properties of the timing model fit to determine how much GW power was absorbed by the fitting procedure. The 95\% confidence upper limit of $7\times10^{-15}$ was based on analysis of data from 17 pulsars timed with either Arecibo or the GBT (one pulsar was timed by both). As with the PPTA and EPTA analyses, the limit was dominated by the residuals
of the two best-timed pulsars in the data set.

\section{PTA Sensitivity}
\label{sec:sens}

The primary goal of the IPTA over the next several years is to increase the sensitivity of our experiment and make the first direct
detection of nanohertz frequency GWs. To reach this goal, it is critical to understand the factors which impact the sensitivity of a PTA to GWs.
These factors include the number of MSPs in the array, ${N}_{\rm MSP}$, the
 average RMS timing residual,
$\sigma_{\rm RMS}$, the total timespan of the observations, $T$, and the cadence (i.e. the inverse of the average spacing in time)
 of observations, $C$.
 The scaling of the minimum detectable characteristic strain with these variables depends on whether or not the
lowest frequencies of the GW power spectrum are above or below the level of white noise present in the data. In the {\it weak} regime, the amplitude of the stochastic background is always below the white noise level, for all GW frequencies. In the {\it strong} regime, the
amplitude of the stochastic background is always above the white noise level. In the weak regime, the signal-to-noise ratio of the stochastic background for amplitude $A$
scales like ${N}_{\rm MSP} c (A/\sigma_{\rm RMS})^2 T^{13/3}$, with a strong dependence on all four quantifies \cite{sejr13}.
In the intermediate regime, where the
lowest frequencies of the stochastic
background are above the white noise level, while the higher frequencies remain below, the ratio scales like ${N}_{\rm MSP} T^{1/2} (A/\sigma_{\rm RMS}\sqrt{c})^{3/13}$ \cite{sejr13}.
 In other words, sensitivity depends only weakly
on the cadence of observations and the average RMS, and much more strongly on the number of pulsars and the total timespan of the observations. Given 
 RMS timing residuals currently achieved by the three PTAs and the predicted levels of the SMBH binary background,
the current PTA experiments are likely in or quickly approaching the {\it intermediate} regime \cite{sejr13}.
{\it Therefore, the most important step to improve sensitivity to a stochastic background of GWs is to increase the number
of MSPs in the array.
}

Note that the above argument assume that all noise intrinsic to the pulsars is white.
While this seems to be true for most MSPs \cite{vbc+09,dfg+13}, it is possible that as timing precisions increase we will
hit a `noise floor' due to  intrinsic spin-down noise with a red spectrum (with predicted power law spectral index of
 $-3$ to $-5$ compared to $-4.3$ for GWs) \cite{cs10}. Therefore,
 spin noise will become more apparent, and have a more detrimental effect on sensitivity, as timespans increase. However, in the case that red noise becomes important (and it has not yet become important for the great majority of IPTA-timed MSPs), the best way to increase
sensitivity is to increase the number of MSPs in the array, so the strategy remains the same. 

The above scaling of sensitivity with observational parameters applies only to stochastic background detection.
For continuous wave or burst source detection, the scaling is quite different, with the number of pulsars being relatively not important
and the MSP timing 
precisions and observing cadence much more important.  Therefore, in order to increase sensitivity to continuous wave and burst sources,
performing high-cadence observations of several of the highest-precision MSPs is optimal \cite{blf11}.

\section{Time to Detection}
\label{sec:detection}

In order to estimate the time to the first detection of GWs with PTAs, models for the expected amplitude of the GW background are necessary.
Several models for the expected level of the stochastic background due to SMBH binaries have been published. The first of
these is an {\it empirical synthesis model} that assumes several different estimates of the redshift-dependent galaxy mass function and of the fraction of close galaxy pairs, coupled with galaxy merger timescales derived from the Millenium simulation \cite{swj+05}, to derive a range of galaxy merger rates \cite{sesana13}. Empirical black hole-host relations are then used to populate merging galaxies with SMBHs. This results in a range of calculated SMBH binary merger rates at redshifts $<$ 1.5.  For each of these merger rates, the GW signal is computed, a large set of estimates of GW background amplitudes are produced, and confidence
intervals for the expected amplitudes in the nHz frequency band are calculated. Using this method, the 3$\sigma$ lower and upper limits for
the predicted SMBH binary background strength are $1\times10^{-16}$ and $4\times10^{-15}$ (see Figure~\ref{fig:limits}). This encompasses the most recent and most sensitive upper limit \cite{src+13}, implying that the chances of detection are non-negligible even over the next several years.

A complementary approach begins with the assumption that the observed evolution of the galaxy mass function can be reproduced under the
assumption of negligible star formation. This {\it merger-driven model} also assumes several different estimates of the redshift-dependent galaxy mass function, but instead of basing the calculation off of  pair fractions, uses the observed mass function to calibrate an analytical model that assumes very massive galaxies evolve primarily due to mergers at redshifts $<$ 1 \cite{mop14}. This model  uses an  updated black hole mass/bulge mass relationship \cite{mm13}, which
incorporates recent measurements of ultra-massive BHs in brightest cluster galaxies. Using this method the 2$\sigma$ lower and upper limits for
the predicted SMBH binary background strength are $1\times10^{-15}$ and $7\times10^{-15}$ (see Figure~\ref{fig:limits}). Much of this range has already been ruled out by the recent PPTA limit \cite{src+13}.
 When comparing this merger-driven model with the empirical synthesis estimates \cite{sesana13}, it is important to note that
the predominant reason for the difference in predicted signal levels is not  due to the merger-driven assumption but due to the merger-driven model's reliance on the updated brightest cluster galaxy data \cite{mm13}. Using these mass measurements in the empirical synthesis model would result in a similarly (and reassuringly!) high predicted SMBH stochastic background level.

Recently, a contrasting approach for estimating the expected SMBH stochastic background level used hydrodynamical cosmological simulations
to conclude that gas accretion remains the dominant source of mass accumulation in almost all of the SMBHs in a cluster \cite{kon+13}. In this
case, the predicted signal is reduced, with 3$\sigma$ lower and upper limits for
the predicted SMBH binary background strength of $6\times10^{-16}$ and $4\times10^{-15}$ (see Figure~\ref{fig:limits}).
These predictions fall in between the empirical synthesis model and the merger-driven model, though they have closer agreement with the merger-driven model.
 
A final approach involves using the Millennium \cite{swj+05}  and Millennium II \cite{bsw+09} dark matter simulations
to create a semi-analytic model in which SMBHs are seeded in every galaxy merger 
remnant at early times and grow primarily by gas accretion triggered by galaxy mergers. This results in 3$\sigma$ lower and upper limits for
the predicted SMBH binary background strength of $5\times10^{-16}$ and $8\times10^{-16}$ (see Figure~\ref{fig:limits}) \cite{src+13}.
 
Overall, despite varying assumptions about the role of mergers and the roles of baryons and dark matter in reproducing the galaxy mass function, in addition to reliance on different observational data, all four of these models are in
reasonably good agreement and indicate that a GW detection through PTAs is possible within the decade.
As upper limits become more stringent with time, the phase space of primarily merger-drive models will continue to become ruled out, and the need for other evolutionary processes such as accretion will become clearer. In any case, it is almost certain
that all current models are too simplistic, showing that actual GW detection and observations in the nanohertz regime
 has enormous potential to inform our understanding of
galaxy evolution.

\begin{figure*}
\centering\includegraphics[width=1.0\textwidth]{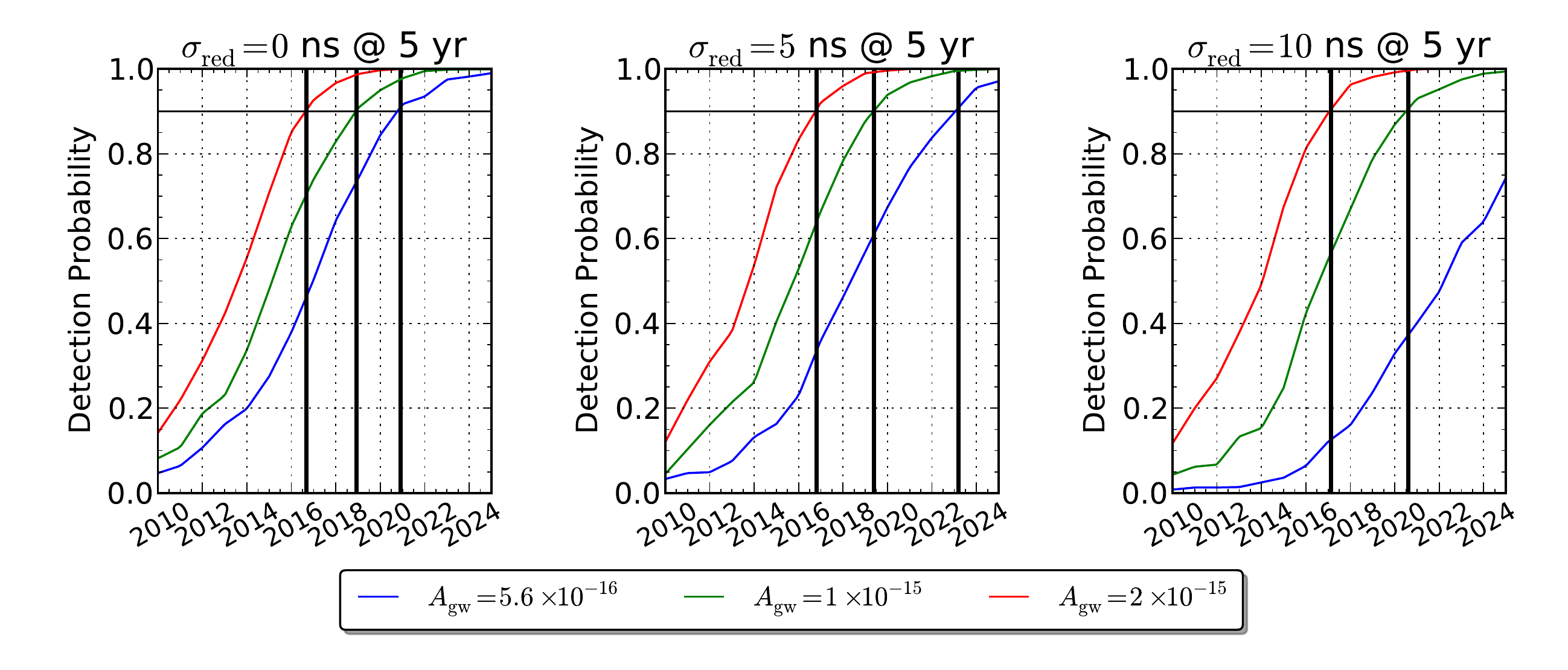}
\caption{Detection probability (at 90\% confidence) vs time in years for the NANOGrav PTA for three different background amplitudes that span the range discussed in the empirical synthesis model \cite{sesana13,sejr13}.
 This assumes 35 MSPs being timed in 2013, increasing by three per year and an average RMS of 200~ns (roughly that observed for the NANOGrav pulsars). These plots show the expectations for three different values of red spin noise that induce an RMS of
0, 5, and 10 ns at 5 yrs. In the most optimistic cases, detection is possible by 2016 and in all but the most pessimistic case, a detection will be made by 2022. Credit: Xavier Siemens.
}
\label{fig:timetodetection}
\end{figure*}

The
scaling relations discussed in Section~\ref{sec:sens} could be used to make
robust time-to-detection estimates assuming the properties of the canonical PTAs.
These estimates have not yet been made for the entire IPTA data set. However, realistic simulations have been carried out
assuming the properties of 
the NANOGrav PTA (see Figure~\ref{fig:timetodetection}). These  show that for all 
 but the most pessimistic scenario (i.e. lowest amplitude and largest intrinsic spin
noise), a detection will occur by 2023, and could occur as early as 2016.
While we have not yet quantified the sensitivity of the full IPTA data set, it will certainly be more sensitive than the NANOGrav data alone, effectively making these estimates upper limits on the time to detection for the IPTA, assuming that the
range of backgrounds predicted by the empirical synthesis model \cite{sesana13} is accurate.

\begin{figure*}
\centering\includegraphics[width=0.7\textwidth]{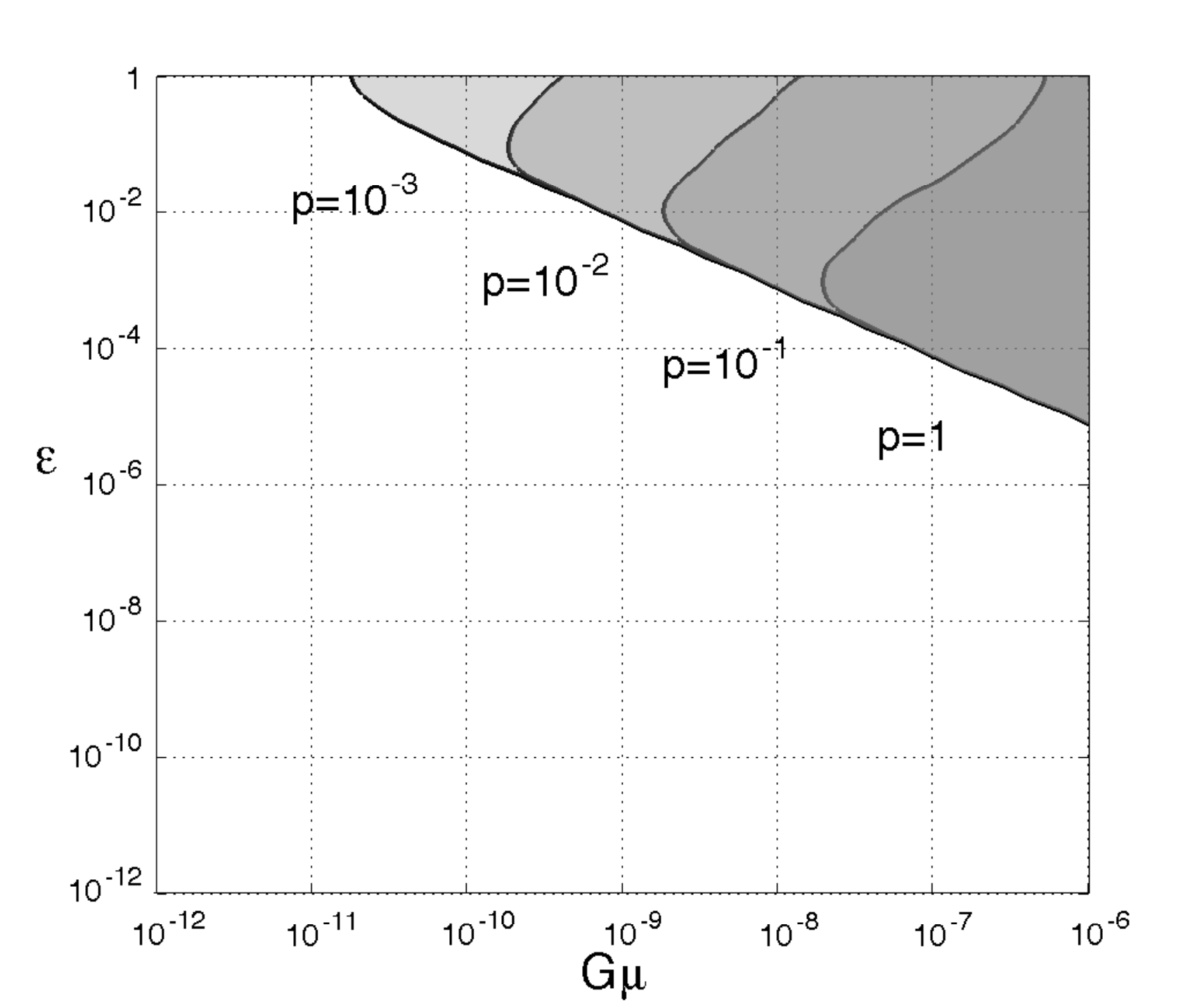}
\caption{
Cosmic string parameter space constraints from the NANOGrav upper limit
measurement, in the `small loop' case, where cosmic string loop sizes are set by gravitational back reaction \cite{dfg+13}. The shaded areas show
regions of string tension ($G\mu$) and loop size ($\epsilon$) that are ruled out
by the measurements. These are shown  for various values of the reconnection probability
($p$).
Reprinted with permission of P. Demorest.
}
\label{fig:strings}
\end{figure*}

The estimates in Figure~\ref{fig:timetodetection} rely on SMBH binaries as the source of the stochastic GW background. The predicted
amplitude of the stochastic background due to cosmic strings depends sensitively on parameters such as loop size, reconnection probability,
and string tension, resulting in a wide range of predicted amplitudes \cite{sbs13,dfg+13}. Current PTA upper limits already rule out some parameter space
for cosmic strings (see Figure~\ref{fig:strings}); these constraints will improve dramatically over the next several years.
The GW stochastic background amplitude expected due to relic GW radiation is very uncertain, depending sensitively on the assumed spectral index and on the
tensor-to-scalar ratio, $r$, which sets the initial amplitude.
Current limits from PTAs already show that the spectral index of the background must be less than $\sim-0.8$ for expected values of $r$ \cite{tong14}.  
The constraints set are not yet as stringent as those in the LIGO/VIRGO band, however, and detection of the relic GW background with
PTAs may not be possible for decades \cite{tong14}.
Indeed, if it lies far below the backgrounds from SMBH binaries or cosmic strings, it will never be detectable.

Though possible,
it is unlikely that the first GW detection in the nanohertz band will arise from a continuous wave search \cite{rws+14}. It
is more likely that the first GW detection may arise from a very anisotropic GW background \cite{sesana13_2}, making the 
developing of algorithms that are sensitive to a variety of correlation curves critical \cite{msmv13}.
It is likely that {\it all} PTA-detectable SMBH binaries will have an identifiable host galaxy and that 30\% may have electromagnetically active SMBHs  \cite{sarah13}. It is therefore possible that the first GW detection with PTAs will arise from a search for GWs from an electromagnetically identified candidate, much like the case of 3C66B \cite{jllw04}. In this case, a PTA detection will allow measurement of
binary parameters which supplement electromagnetic information, and will lead to joint GW and radio/x-ray observations.

\section{The Future of Low Frequency Gravitational Wave Astrophysics}

One of the foremost goals of fundamental physics is the direct detection of GWs, which will enable a new era of GW astrophysics.
In support of that, the primary goal of the PPTA, EPTA, and NANOGrav, and of the broader IPTA consortium, over the past several years has been
{\it building and characterizing} a sensitive GW detector. This has involved searches for MSPs with  the world's largest radio telescopes,
characterization of the properties of the MSPs, and determination of the optimal techniques for taking, processing, storing, and sharing
timing
data. With the first PTA GW detection, the focus of the community's work will shift from building and characterization to {\it GW source
 characterization}. We expect this shift to happen within the next decade, and possibly within the next several years, 
given our current understanding of source populations and the current and projected sensitivities of the IPTA. A detection of
a stochastic background of GWs will enable constraints on source populations and will foster the development of more accurate models for
the formation and evolution of galaxies. It will also allow sensitive tests of general relativity as other polarization modes would result in  slightly different cross-correlation signatures. 
Detection of continuous wave sources and bursts will allow characterization of individual SMBH binaries which, coupled with electromagnetic observations, will revolutionize our views of galaxy formation and evolution and test our assumptions about general relativity. As we move further into the future, with the advent of telescopes such as the FAST, MeerKAT, and, ultimately, the Square Kilometer Array (SKA), hundreds of high-precision MSPs will be timed, with precisely measured distances, allowing us not only to characterize but localize GW sources, leading to a true Galactic-scale GW observatory.

\begin{acknowledgements}
Thank you to all of my colleagues in NANOGrav and the IPTA for many conversations that have improved this paper. The comments from the anonymous referee were also immensely helpful. The author is supported through NSF PIRE award \#0956296 for this work.
\end{acknowledgements}

\bibliographystyle{spphys}       
\bibliography{bib,refother}   

\end{document}